# BRAIN TUMOR CLASSIFICATION FROM MRI IMAGES USING MACHINE LEARNING

**Vidhyapriya Ranganathan\* Celshiya Udaiyar\*, Jaisree Jayanth\*, Meghaa P V\*, Srija B\*, Uthra S\***
\* Department of Biomedical Engineering, PSG College of Technology, India

## Abstract:

Brain tumor is a life-threatening problem and hampers the normal functioning of the human body. The average five-year relative survival rate for malignant brain tumors is 35.6 percent. For proper diagnosis and efficient treatment planning, it is necessary to detect the brain tumor in early stages. Due to advancement in medical imaging technology, the brain images are taken in different modalities. The ability to extract relevant characteristics from magnetic resonance imaging (MRI) scans is a crucial step for brain tumor classifiers. Several studies have proposed various strategies to extract relevant features from different modalities of MRI to predict the growth of abnormal tumors. Most techniques used conventional methods of image processing for feature extraction and machine learning for classification. More recently, the use of deep learning algorithms in medical imaging has resulted in significant improvements in the classification and diagnosis of brain tumors. Since tumors are located at different regions of the brain, localizing the tumor and classifying it to a particular category is a challenging task. The objective of this project is to develop a predictive system for brain tumor detection using machine learning(ensembling).

The datasets Brain Tumor MRI Images for Brain Tumor Detection(named as 'Dataset 1') and Brain Tumor Classification (MRI)(named as 'Dataset 2') are collected and are preprocessed. The datasets are trained in various machine learning models and their accuracies are evaluated. Dataset1 is trained in models DenseNet, ResNet, EfficientNet and VGG-16 and Dataset2 is trained in DenseNet, ResNet, EfficientNet and ViT models. Models trained with Dataset1 will predict the presence or absence of a tumor whereas the models trained with Dataset 2 will classify the types of tumors.

VGG-16 produced higher accuracy compared to all other models trained with Dataset1 and DenseNet produced higher accuracy with Dataset2. These two models are ensembled to further increase the accuracy of multiclass brain tumor classification

# CHAPTER 1

# INTRODUCTION

The brain is one of the largest and most complex organs in the human body. Figure 1.1. Shows an image of the Brain. It is made up of more than 100 billion nerves that communicate in trillions of connections called synapses. The brain is made up of specialized areas that are mentioned below:

- The cortex is the outermost layer of brain cells. Thinking and voluntary movements begin in the cortex.
- The brain stem is between the spinal cord and the rest of the brain. Basic functions like breathing and sleep are controlled here.
- The basal ganglia are a cluster of structures in the center of the brain. They coordinate messages between multiple areas of the brain.
- The cerebellum is at the base and the back of the brain. It is responsible for coordination and balance.

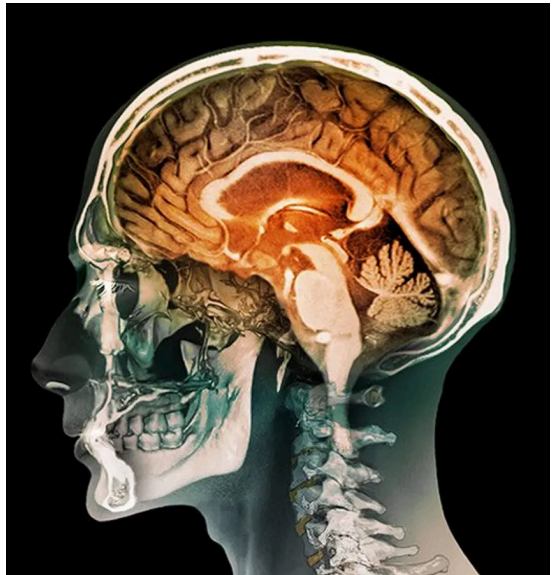

*Fig. 1.1. Brain Image*

**Brain Conditions**

1. **Headache:** There are many types of headaches; some are serious but most are not and are generally treated with painkillers.
2. **Stroke :** Blood flow and oxygen are suddenly interrupted to an area of brain tissue, which then dies. A blood clot, or bleeding in the brain, are the cause of most strokes.

3. **Brain tumor:** Any abnormal growth of tissue inside the brain. Whether malignant (cancer) or benign, brain tumors usually cause problems by the pressure they exert on the normal brain.
4. **Glioblastoma:** An aggressive, malignant brain tumor (cancer). Brain glioblastomas progress rapidly and are very difficult to cure.
5. **Parkinson's disease:** Nerves in a central area of the brain degenerate slowly, causing problems with movement and coordination. A tremor of the hands is a common early sign.

## 1.1 BRAIN TUMOR

A tumor is a mass or lump of cells that can develop in any part of the body. Tumors can be benign or malignant. Benign tumors are non-cancerous growths that do not spread to other parts of the body, while malignant tumors are cancerous growths that have the ability to spread to other parts of the body (a process known as metastasis) and can be life-threatening.Figure 1.2 shows the MRI Image of a tumor.

A tumor and a cyst are both abnormal growths that can develop in the body, but they differ in several ways.

Tumors can be further classified based on their cellular origin and histopathological characteristics. For example, tumors that arise from the cells that make up the nervous system are called neurogenic tumors, while tumors that arise from the cells that make up the glandular tissues are called adenomas.

On the other hand, a cyst is a closed sac-like structure that can be filled with fluid, gas, or semi-solid material. Cysts can develop in any part of the body, including the brain, ovaries, kidneys, liver, and skin. Cysts can be either benign or malignant, and they can vary in size from a few millimeters to several centimeters. Figure 1.3. shows the MRI Image of a cyst.

The symptoms and treatment of tumors and cysts can vary depending on their location, size, and histopathological characteristics. In some cases, surgery may be necessary to remove the growth, while in other cases, the growth may be monitored or treated with medication.

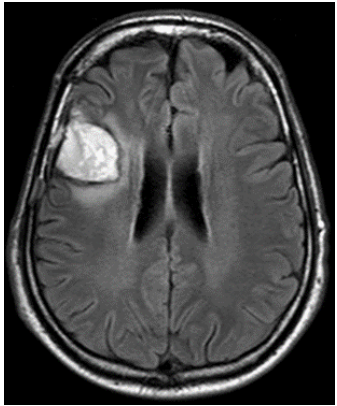
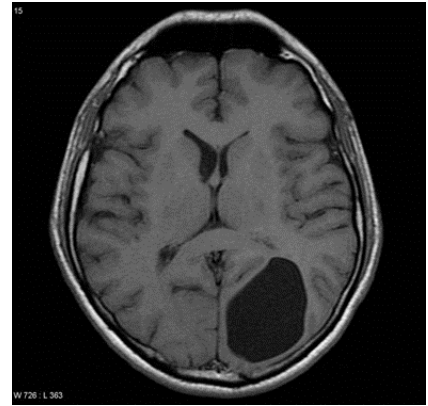

*Fig.1.2.  Tumor*	*Fig.1.3.  Cyst*

### 1.1.1. Types of Brain Tumor:

The terms primary and metastatic describe where the tumor has originated, and brain tumors are generally classified as one or the other. Primary brain tumors arise from the brain or spinal cord while metastatic brain tumors, also termed secondary brain tumors, arise from other tissue and have spread to the brain. This is the most basic form of classifying brain tumors, but yields great insight into the characteristics of these complex growths, and how they might be treated.

**a.    Primary Brain Tumors:**

Primary brain tumors are tumors that originate in the brain or surrounding tissues, such as the meninges or cranial nerves. These tumors are different from secondary brain tumors, which originate in other parts of the body and metastasize (spread) to the brain.

Primary brain tumors can be further classified into several subtypes based on their cellular origins and histopathological characteristics. Some of the most common primary brain tumor subtypes include gliomas, meningiomas, pituitary adenomas, ependymomas, and schwannomas.

- **Glioma:**

Gliomas are the most common type of primary brain tumor and arise from the glial cells in the brain. These tumors can be further classified based on their cellular and molecular characteristics. The classification is based on whether they have mutations in the isocitrate dehydrogenase (IDH) gene or co-deletion of chromosomes 1p and 19q. Gliomas can range from low-grade (slow-growing) to high-grade (fast-growing) tumors, with the most aggressive subtype being glioblastoma. Figure 1.4 shows an MRI image of Glioma.

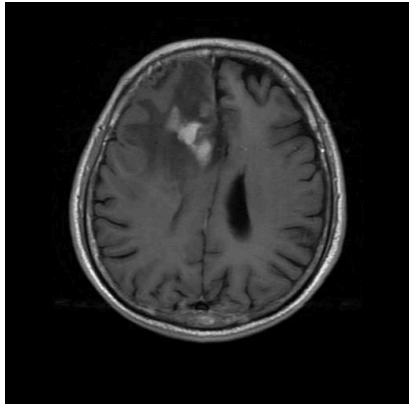
*Fig. 1.4. Glioma*

- **Meningioma:**

Meningiomas are tumors that arise from the meninges, the protective layers of tissue that surround the brain and spinal cord. They are typically slow-growing and can often be surgically removed. Fig. 1.5. shows an MRI image of Meningioma.

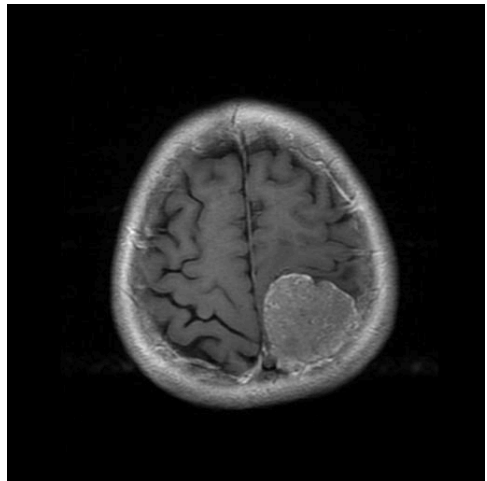
*Fig 1.5. Meningioma*

- **Pituitary adenoma:**

Pituitary adenomas are tumors that arise from the pituitary gland, a small gland located at the base of the brain that produces hormones. These tumors can cause hormonal imbalances and are often treated with surgery, radiation therapy, or medication. Figure 1.6. shows the MRI image of Pituitary adenoma.

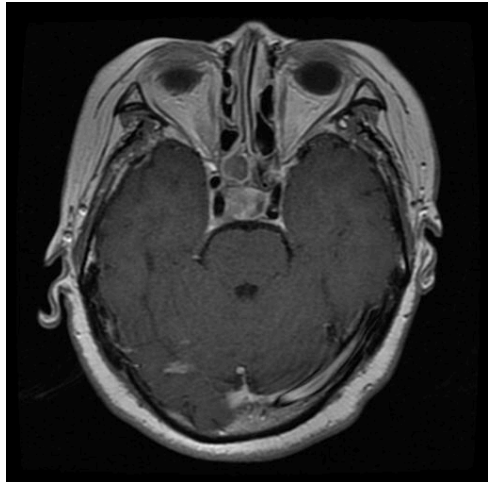
*Fig 1.6. Pituitary adenoma*

- **Ependymoma:**

Ependymomas arise from the cells lining the ventricles (fluid-filled spaces) in the brain or the central canal of the spinal cord. Figure 1.7. Shows the MRI image of Ependymoma. These tumors can be slow-growing or fast-growing, depending on their location and histopathological characteristics. Schwannomas are tumors that arise from the Schwann cells that wrap around nerve fibers in the brain and spinal cord. They are often slow-growing and can often be surgically removed.

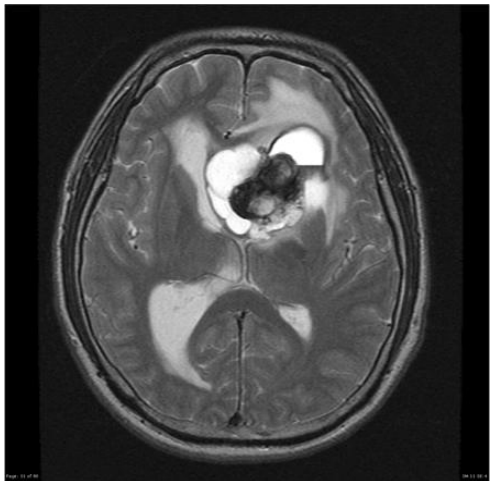
*Fig.1.7. Ependymoma*

b. **Secondary Brain Tumors:**

Secondary brain tumors, also known as metastatic brain tumors, are tumors that originate in other parts of the body and spread to the brain. These tumors are much more common than primary brain tumors and can occur in up to 30% of patients with cancer.

The most common types of cancer that can metastasize to the brain include lung cancer, breast cancer, melanoma, and colon cancer. The spread of cancer to the brain can occur through the bloodstream or through direct extension from nearby tissues.

The symptoms of secondary brain tumors can vary depending on the location and size of the tumor, but they often include headache, seizures, weakness or numbness in the limbs, changes in vision or hearing, and cognitive changes.

Treatment options for secondary brain tumors typically involve a combination of surgery, radiation therapy, and chemotherapy. The prognosis for patients with secondary brain tumors depends on the extent of the disease and the underlying cancer type, but it is generally worse than for patients with primary brain tumors.

## 1.2. MAGNETIC RESONANCE IMAGING(MRI):

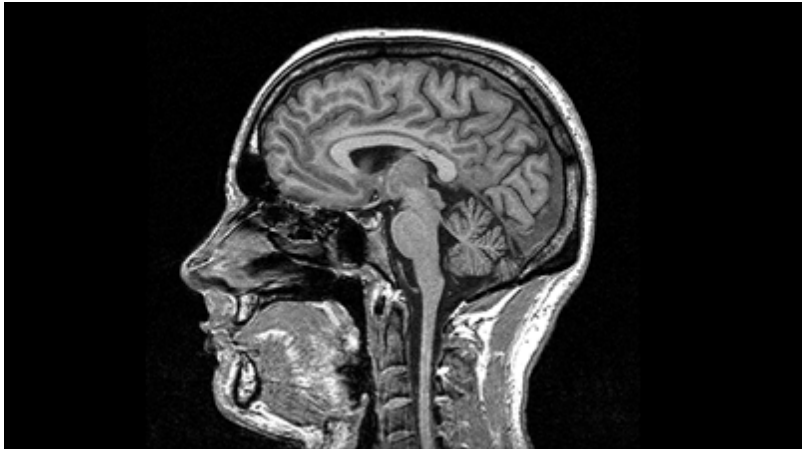

*Fig 1.8. MRI Image*

MRI is a medical imaging technique that uses a magnetic field and radio waves to create detailed images of the internal structures of the body. Figure 1.8. shows an MRI Image. It uses a strong magnetic field and radio waves to generate images of the body's internal structures.

The human body contains hydrogen atoms, which are abundant in water and fat molecules. When the patient is placed inside the strong magnetic field of the MRI machine, the protons (hydrogen atoms) in their body become aligned in the direction of the magnetic field. A brief burst of radio waves is then directed at the protons, causing them to absorb energy and temporarily move out of alignment. This process is known as "excitation".

Once the radio waves are turned off, the protons return to their original state and release the absorbed energy in the form of a radiofrequency signal. This signal is detected by the MRI machine's receiver coils and processed by a computer to create an image of the body's internal structures.

In the case of brain imaging, MRI can be used to create high-resolution images of the brain and its surrounding tissues. MRI brain imaging is a highly effective tool for detecting abnormalities in the brain, including brain tumors. MRI can produce high-resolution images of the brain and surrounding tissues, allowing for a detailed examination of any abnormalities.

When a brain tumor is present, it can appear as a mass or area of abnormal tissue on the MRI images. The location and size of the tumor can be accurately determined by examining the images, and the type of tumor can often be identified based on its appearance.

In addition to detecting the presence of a tumor, MRI can also provide information about the extent of its involvement with surrounding tissues. This is important in determining the best course of treatment for the tumor. For example, if the tumor is located near critical structures such as blood vessels or nerves, surgery may be more difficult or risky. In such cases, MRI can help in planning the surgical approach and determining the optimal time for surgery.

MRI can also be used to monitor the progression of a brain tumor over time. By comparing images taken at different points in time, doctors can evaluate the effectiveness of treatment and determine if the tumor is growing or shrinking.

Overall, MRI brain imaging is an invaluable tool in the detection and management of brain tumors. Its high-resolution images and ability to provide detailed information about the location, size, and extent of a tumor make it an essential part of the diagnostic and treatment process for patients with brain tumors.

MRI brain imaging has several advantages over other imaging techniques in detecting brain tumors:

- **High-resolution images:** MRI can produce very high-resolution images of the brain, which allows for a detailed examination of the tumor and surrounding tissues. This can help in identifying the type of tumor, its location, and its relationship with nearby structures, all of which are important factors in determining the most effective treatment plan.

- **Non-invasive:** MRI is a non-invasive imaging technique, which means that it does not require any surgical incisions or injections. This reduces the risk of complications and makes the procedure more comfortable for the patient.

- **Multi-planar imaging:** MRI can produce images in multiple planes (e.g., sagittal, coronal, and axial), which allows for a more comprehensive evaluation of the tumor and its relationship with surrounding structures.

- **No ionizing radiation:** MRI does not use ionizing radiation, which is a type of radiation that can damage cells and increase the risk of cancer. This makes MRI a safer imaging option, particularly for patients who require multiple imaging studies over time.

- **Contrast-enhanced imaging:** MRI can be performed with the use of contrast agents, which are substances that can help to highlight certain structures or abnormalities in the brain. This can help in identifying small or subtle tumors that may not be visible on non-contrast images.

## 1.3. MRI IMAGING SEQUENCES

T1-weighted (T1W) and T2-weighted (T2W) images are two of the most commonly used MRI sequences for clinical and research purposes.

T1-weighted images are acquired with a short repetition time (TR) and a short echo time (TE), resulting in high signal intensity for tissues with short T1 relaxation time (e.g., fat) and low signal intensity for tissues with long T1 relaxation time (e.g., water). T1W images are useful for imaging anatomical details, such as the shape and size of organs, and for detecting lesions that have low water content, such as hemorrhage and calcifications. T1W images are also used for imaging the brain, where they can differentiate gray matter, white matter, and cerebrospinal fluid (CSF).

On the other hand, T2-weighted images are acquired with a long TR and a long TE, resulting in high signal intensity for tissues with long T2 relaxation time (e.g., water) and low signal intensity for tissues with short T2 relaxation time (e.g., fat). T2W images are useful for detecting pathology that alters the water content of tissues, such as edema, inflammation, and tumors. T2W images are also used for imaging the brain, where they can differentiate gray matter, white matter, and CSF based on their T2 relaxation times.

Figure 1.9. represents the images of T1 and T2 weighted MRI Images.

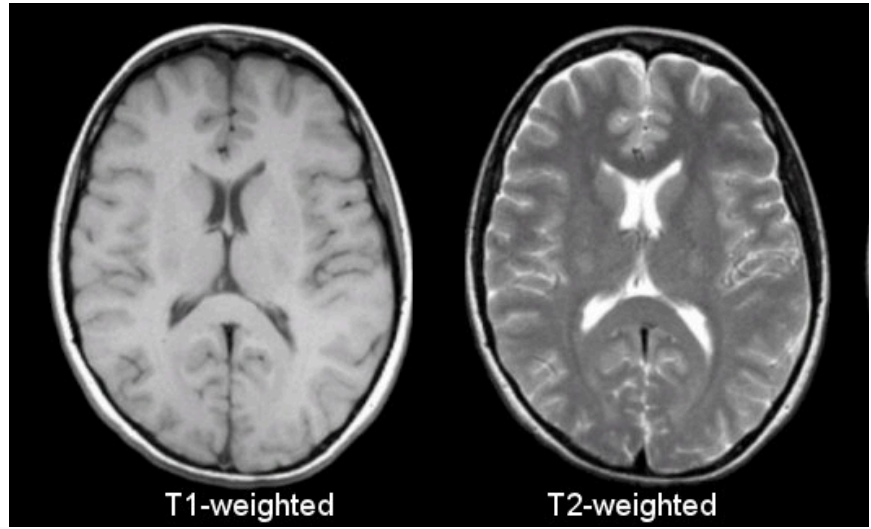

*Fig. 1.9. T1 and T2 weighted MRI images*

In general, T1- and T2-weighted images can be easily differentiated by looking at the CSF. CSF is dark on T1-weighted imaging and bright on T2-weighted imaging.

The main difference between T1W and T2W images is their contrast characteristics, which are determined by the relaxation times of tissues. T1W images have low signal intensity for tissues with long T1 relaxation time, which includes most soft tissues and fluids. However, T2W images have high signal intensity for tissues with long T2 relaxation time, which includes most fluids, such as CSF, blood, and edema. Therefore, T1W images are best for imaging structures with high fat content, while T2W images are best for imaging structures with high water content.

In clinical practice, T1W and T2W images are often acquired together as part of a standard MRI protocol. This allows for complementary information to be obtained, which can aid in diagnosis and treatment planning. For example, in brain imaging, T1W images are used to detect anatomical changes, while T2W images are used to detect lesions and abnormalities in the brain parenchyma. Table 1.1. shows the comparison between T1 and T2 weighted images.

*Table 1.1. Comparison of T1 vs T2 weighted images*

| Tissue | T1-Weighted | T2-Weighted |
|---|---|---|
| CSF | Dark | Bright |
| White matter | Light | Dark Gray |
| Cortex | Gray | Light Gray |
| Fat (within bone marrow) | Bright | Light |
| Inflammation (infection, demyelination) | Dark | Bright |

## 1.4. MACHINE LEARNING

Machine learning is an important component of data science. Through the use of statistical methods, algorithms are trained to make classifications or predictions and to uncover key insights in data mining projects.

### a. Machine learning vs Deep learning vs Neural networks:

Since deep learning and machine learning tend to be used interchangeably, it's worth noting the nuances between the two. Machine learning, deep learning, and neural networks are all sub-fields of artificial intelligence. However, neural networks is actually a sub-field of machine learning, and deep learning is a sub-field of neural networks.

### b. How Machine Learning Works

1. **A Decision Process:** In general, machine learning algorithms are used to make a prediction or classification. Based on some input data, which can be labeled or unlabeled, your algorithm will produce an estimate about a pattern in the data.
2. **An Error Function:** An error function evaluates the prediction of the model. If there are known examples, an error function can make a comparison to assess the accuracy of the model.
3. **A Model Optimization Process:** If the model can fit better to the data points in the training set, then weights are adjusted to reduce the discrepancy between the known example and the model estimate. The algorithm will repeat this "evaluate and optimize" process, updating weights autonomously until a threshold of accuracy has been met.

**c. Key elements of Machine Learning**

There are three main elements to every machine learning algorithm, and they include:

- Representation: what the model looks like; how knowledge is represented
- Evaluation: how good models are differentiated; how programs are evaluated
- Optimization: the process for finding good models; how programs are generated

**d. Image Classification using Machine Learning:**

Image classification is a supervised learning problem: define a set of target classes (objects to identify in images), and train a model to recognize them using labeled example photos. Early computer vision models relied on raw pixel data as the input to the model. The position of the object, background behind the object, ambient lighting, camera angle, and camera focus all can produce fluctuation in raw pixel data; these differences are significant enough that they cannot be corrected for by taking weighted averages of pixel RGB values.

## 1.5. PROBLEM STATEMENT

Distinguishing brain tumors from other cysts or tumor like growths from MRI images of the brain using Machine learning models to detect and diagnose at the earlier stages

Types of Brain tumors based on the location of the tumors in and around the brain region are

- Glioma - Begins in the brain or spinal cord
- Meningioma - Arises from the membranes that surround brain and spinal cord (meninges)
- Pituitary Tumor - Unusual growths that develop in the pituitary gland

## 1.6. OBJECTIVES

- To identify tumors from cysts that look like tumors from MRI images using machine learning.
- To classify the tumors using image classification models like DenseNet, Efficient Net, ResNet and Vision Transformer.
- To ensemble a model that identifies tumors and classifies the tumors.

# CHAPTER 2

# LITERATURE SURVEY

This chapter presents the literature survey on Brain Tumor classification of MRI images using Machine learning techniques. Literature survey includes all types of published literature. This survey consists of all existing literature related to the proposed work.

In the paper **'MRI-Based Brain Tumor Classification Using Ensemble of Deep Features and Machine Learning Classifiers'** by Jaeyong Kang ,Zahid Ullah and Jeonghwan Gwak, their main objective is to classify the brain tumors using the ensemble of deep features from pre-trained deep CNN's with ML classifiers from three different MRI dataset. The concept of transfer learning and several other pre-trained CNN's were adopted to extract features from brain MRI. The extracted deep features are then evaluated by several ML classifiers and which resulted in better classification. The outcome of this method achieved an enormous improvement with prominent accuracy for the classification of brain tumors.

The paper **'Classification of Brain Tumor Types Based On MRI Images Using Mobilenet'** by Tsamara Hanifa Arfan, Mardhiya Hayaty and Arifiyanto Hadinegoro focuses on classifying the brain tumors from MRI images using deep learning techniques and here MobileNet is used. Mobile Network is a CNN model with high accuracy and less computation. Hence this paper proposed a solution for brain Tumor classification by using the state-of-the-art MobileNet V2140 x 224 architecture in improving the classification accuracy which finally achieved the accuracy of 94% with better improvement and the algorithm outperforms itself in classifying the data with much lesser computation loss.

In the paper **'MRI Brain Tumor Classification Using a Hybrid VGG16-NADE Model'** by Saran Raj Sowrirajan, Surendiran BalasubramanianRaja and Soosaimarian Peter Raj, a hybrid VGG16_NADE model has been proposed for brain tumor classification from MRI images dataset considering the T1 weighted images. The VGG16 pre-trained CNN consists of 16 layers that extract detailed features from MRI images and a neural autoregressive density estimator (NADE) removes redundant brain images and smoothens the tumor border. The basic network of this model is inspired by VGG16 and experiments are conducted on an open-access series of imaging studies (OASIS). The experiments show that the proposed model outperforms the state-of-the-art models in terms of accuracy.

In the paper **'A Transfer Learning approach for AI-based classification of brain tumors'** by Rajat Mehrotra ,M.A. Ansari, Rajeev Agrawal b and R.S. Anand, an AI-based classification of brain tumor into Malignant and Benign using Deep Learning Algorithms are proposed for the classifying types of brain tumors utilizing openly accessible datasets. The relevance of Artificial Intelligence (AI) in the form of Deep Learning (DL) in the area of medical imaging has paved the path to extraordinary developments in categorizing and detecting intricate pathological conditions, like a brain tumor. It is evident that transfer learning through AlexNet gives the utmost performance.

In the paper **'DenseNet For Brain Tumor classification of MRI Images'** by AbhijeetA.Gajre, Omkar S. Khaladkar and Abhijit J. Patil's, the main objective is to present a detailed fundamental analysis of the research and findings performed to detect and classify brain tumors through MRI images. Deep learning algorithms had proven to be the better version for classification of various medical imaging. The dataset used has a total of around 3064 T1- weighted contrast-enhanced images from 233 patients with three kinds of brain tumor. A comparative analysis is made amongst various deep learning techniques and is classified into Meningioma, Glioma and Pituitary Tumor.

In the paper **'Enhanced Watershed Segmentation Algorithm-Based Modified ResNet50 Model for Brain Tumor Detection'** by Arpit Kumar Sharma, Amita Nandal and Arvind Dhaka, the authors have proposed a novel technique to detect brain tumor with the help of enhanced watershed modeling integrated with a modified ResNet50 architecture. The model uses the ResNet50 model with a modified layer architecture including five convolutional layers and three fully connected layers and can retain the optimal computational efficiency with high-dimensional deep features. This work obtains a comprised feature set by retrieving the diverse deep features from the ResNet50 deep learning model and feeds them as input to the classifier. The good performing capability of the proposed model is achieved by using hybrid features of ResNet50.

The paper '**Brain Tumor classification using Dense Efficient-Ne**t' by Dillip Ranjan Nayak, Neelamadhab Padhy  and Pradeep Kumar Mallick, proposed a CNN-based dense EfficientNet using min-max normalization to classify 3260 T1-weighted contrast-enhanced brain magnetic resonance images into four categories (glioma, meningioma, pituitary, and no tumor). The dense CNN model can accurately categorize a limited database of pictures and the approach provides an exceptional overall performance. With high accuracy and a favorable F1 score, the newly designed EfficientNet CNN architecture can be a useful decision-making tool for classification and in the study of brain tumor diagnostic tests.

The paper **'Brain Tumor Classification from MRI using Vision Transformers Ensembling'** by Sudhakar Tummala focuses on Automated classification of brain tumors by Vision Transformer (ViT) based deep neural network architectures based transfer learning approach for achieving good tumor classification accuracy. A brain tumor dataset from figshare consisting of 3064 T1-weighted contrast-enhanced (CE) magnetic resonance imaging (MRI) slices with meningioma, glioma, and pituitary tumor was used for cross-validation and testing of ensembled ViT models ability for 3-class classification task. An inference table for Literature Survey is presented in the Table 2.1

In the paper '**Role of Ensemble Deep Learning for Brain Tumor Classification in Multiple Magnetic Resonance Imaging Sequence Data'** by Gopal S. Tandel, Ashish Tiwari, Omprakash G. Kakde, Neha Gupta, Luca Saba, Jasjit S. Suri, a method to maximize the classification ability between low-grade versus high-grade glioma is proposed. Five well-established convolutional neural networks were adopted for tumor classification. An ensemble algorithm was proposed using the majority vote of five deep learning (DL) models to produce more consistent and improved results than any individual model. The proposed ensemble algorithm is based on the probabilistic prediction for desired labels and the majority voting (MajVot) mechanism of five models. The votes of each class label were calculated using the projected probability of each CNN for each test sample. The proposed ensemble algorithm (MajVot) showed significant improvements in the average accuracy

*Table 2.1. Inference Table for Literature survey*

| S.No | Title | Author | Journal | Year | Inference |
|---|---|---|---|---|---|
| 1. | MRI-Based Brain Tumor Classification Using Ensemble of Deep Features and Machine Learning Classifiers | Jaeyong Kang, Zahid Ullah and Jeonghwan Gwak | Molecular Diversity Preservation International (MDPI) | 2021 | Classification of brain tumors using the ensemble of deep features from pre-trained deep CNN's with ML classifiers from three different MRI dataset. |
| 2. | Classification of Brain Tumors Types Based On MRI Images Using Mobilenet | Tsamara Hanifa Arfan, Mardhiya Hayaty and Arifiyanto Hadinegoro | Journal of the Institute of Electrical and Electronics Engineers | 2021 | Brain tumor classification of MRI images using the state-of-the-art MobileNet V2140 x 224 architecture for improving the classification accuracy. |
| 3. | MRI Brain Tumor Classification Using a Hybrid VGG16-NADE Model | Saran Sowrirajan, Surendiran Balasubramanian and Soosaimarian Peter Raj | Brazilian Archives of Biology and Technology (BABT) | 2022 | A hybrid VGG16_NADE(neural autoregressive density estimator) removes redundant brain images and smoothens the tumor border. |
| 4. | A Transfer Learning approach for AI-based classification of brain tumors | Rajat Mehrotra, M.A. Ansari, Rajeev Agrawal b and R.S. Anand | Elsevier (Machine Learning with Applications 2) | 2020 | AI-based classification of brain tumor into Malignant and Benign using DL Algorithms like ALexNet |
| 5. | DenseNet For Brain Tumor classification of MRI Images | AbhijeetA.Gajre, Omkar S. Khaladkar and Abhijit J. Patil | International Journal of Innovative Science and Research Technology | 2022 | A fundamental analysis of the research and findings performed to detect and classify brain tumors through MRI image using Deep learning algorithms |

| S.No | Title | Author | Journal | Year | Inference |
|---|---|---|---|---|---|
| 6. | Enhanced Watershed Segmentation Algorithm-Based Modified ResNet50 Model for Brain Tumor Detection | Arpit Kumar Sharma, Amita Nandal and Arvind Dhaka | BioMed Research International | 2022 | A technique to detect brain tumor using enhanced watershed modeling integrated with a modified ResNet50 architecture with the retrieved features for better accuracy |
| 7. | Brain Tumor classification using Dense Efficient-Net | Dillip Ranjan Nayak, Neelamadhab Padhy and Pradeep Kumar Mallick | Molecular Diversity Preservation International (MDPI) | 2022 | A CNN-based dense EfficientNet using min-max normalization to classify MRI images into glioma, meningioma, pituitary, and no tumor |
| 8. | Brain Tumor Classification from MRI using Vision Transformers Ensembling | Sudhakar Tummala | Research Square | 2022 | A Vision Transformer (ViT) based deep neural network architectures based transfer learning approach for achieving good tumor classification accuracy. |
| 9. | Role of Ensemble Deep Learning for Brain Tumor Classification in Multiple Magnetic Resonance Imaging Sequence Data | Gopal S. Tandel, Ashish Tiwari, Omprakash G. Kakde, Neha Gupta, Luca Saba, Jasjit S. Suri | Molecular Diversity Preservation International (MDPI) | 2022 | An ensemble algorithm was proposed using the majority vote of several deep learning (DL)models to produce more consistent and improved results than any individual model. The votes of each class label were calculated using the projected probability of each CNN for each test sample. |

**INFERENCE**

Different types of medical imaging technologies such as CT, PET and MRI are used to observe the internal parts of the human body. Among all these imaging modalities, MRI is considered most preferable as it is the only modality that offers valuable information in 2D and 3D formats about brain tumor type, size, shape, and position. However, manually reviewing these images is time-consuming, hectic, and even prone to error. To overcome this problem, several efforts have been made to develop a highly accurate and robust solution for the automatic diagnosis and classification of brain tumors.

The use of deep learning algorithms in medical imaging has resulted in significant improvements. The deep learning-based techniques automatically extract meaningful features which offer significantly better performance.

Various ML models are developed to identify the normal and abnormal brain MRI images. To investigate the benefits of combining different ML models, an ensemble model for the MRI-based brain tumor classification task has been developed.

# CHAPTER 3

# METHODOLOGY

The proposed methodology is designed for Brain Tumor Classification from MRI Images using Machine Learning. The details about the dataset used in the analysis is discussed in this chapter.

## 3.1. DATASETS

Radiologists use magnetic resonance imaging (MRI) as a medical imaging technique to produce images of the anatomy and physiological functions of the body. By combining radio waves, strong magnetic fields, and magnetic field gradients, MRI scanners can create images of the body's organs. Our bodies' MRIs are more frequently used to find brain tumors. This project will use MRI images to find and classify brain tumors.

A series of tests were conducted on two different publically accessible brain MRI datasets for the purposes of classifying brain tumors. (i)The first dataset was taken from the Kaggle website: Brain Tumor MRI Images for Brain Tumor Detection by Navoneel Chakrabarty[1]. For our simplicity, this dataset was named 'Dataset 1'. The total number of samples was 253, of which 98 were benign samples while 155 images were malignant brain tumor samples. Fig. 3.1.a. shows "No" samples (benign), and Fig. 3.1.b. shows "Yes" samples (malignant). The resolution of each image is between 88 x 88 dpi to 300 x 300 dpi, and they are 2-dimension (2D) images.

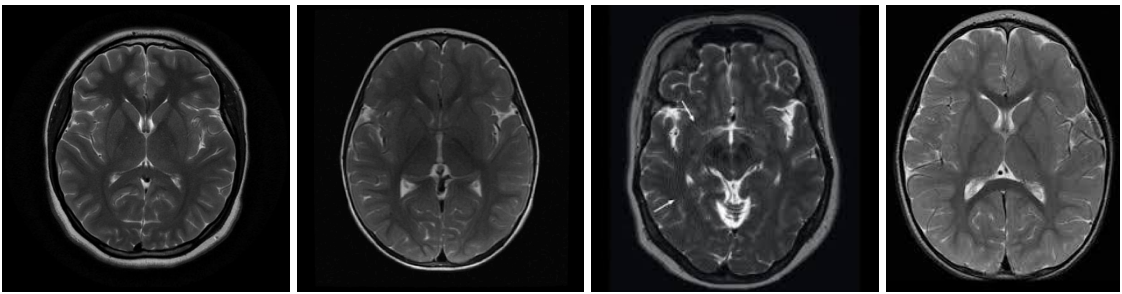
*Fig 3.1.a. Benign images ("No") from Dataset 1*

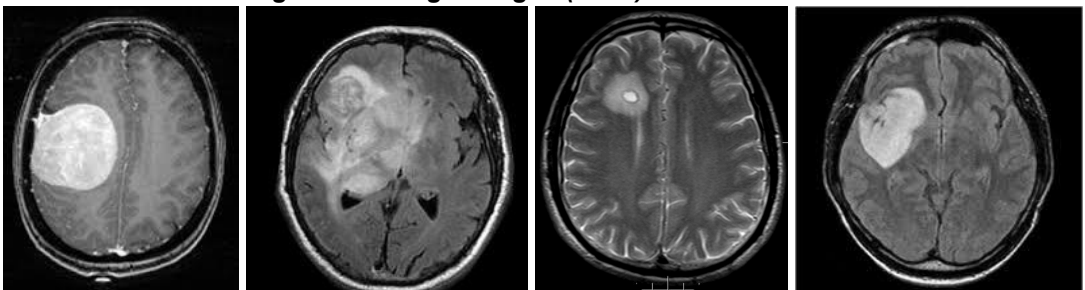
*Fig 3.1.b. Malignant images ("Yes") from Dataset 1*

(ii)The 2nd dataset consists of 3064 T1-weighted images containing three different types of brain tumors: gliomas, meningiomas, and pituitary tumors. In this dataset there is a total of 3064 images where 926 were glioma, 937 were meningioma, 901 were pituitary tumors and 500 were no tumors. The dataset was acquired from the Kaggle website [2] by Sartaj, and this dataset was named 'Dataset 2'. Fig. 3.2.a. shows "glioma_tumor", Fig 3.2.b. shows "meningioma_tumor", Fig 3.2.c. shows "pituitary_tumor" and Fig 3.2.d. shows "no_tumor".

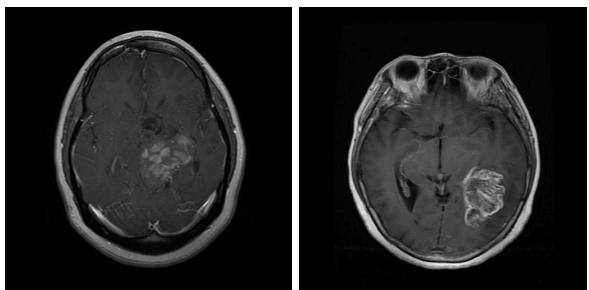
*Fig 3.2.a. Glioma images from dataset 2*

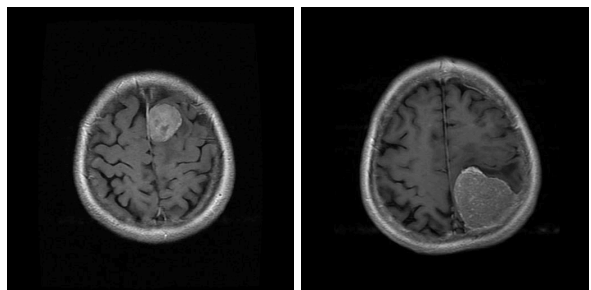
*Fig 3.2.b. Meningioma images from dataset 2*

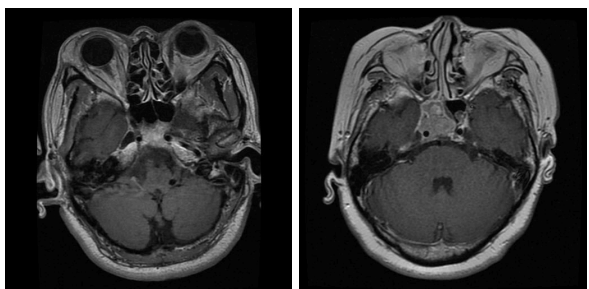
*Fig 3.2.c. Pituitary images from dataset 2*

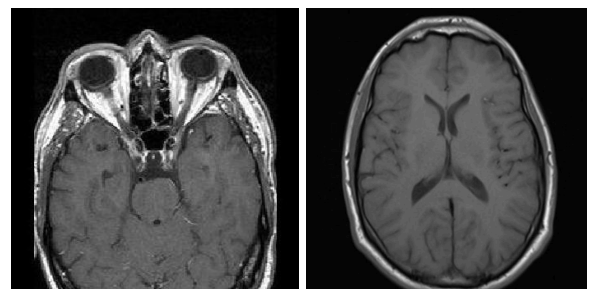
*Fig 3.2.d. No tumor images from dataset 2*

Table 3.1 shows the number of images present across each class of tumor in Dataset 1 and Dataset 2.

*Table 3.1. No. of images present in each class of Tumor in Dataset 1 and Dataset 2*

| DATASET 1 | NO. OF IMAGES | DATASET 2 | NO. OF IMAGES |
|---|---|---|---|
| Tumor | 253 | Glioma | 926 |
| | | Meningioma | 937 |
| | | Pituitary Tumor | 901 |
| No Tumor | 98 | No Tumor | 500 |

## 3.2. DATA PRE-PROCESSING

In order to prepare images for use in model training and inference, a process known as image preprocessing is used. Resizing, orienting, and color adjustments fall under this category. Image augmentation is the process of applying manipulations to images to produce various versions of similar content to provide the model with a wider variety of training examples. A model must consider how an image subject might appear in various scenarios when, for example, randomly changing the rotation, brightness, or scale of an input image.

However, there is a crucial distinction between image augmentation and image preprocessing: while image preprocessing steps are applied to both training and test sets, image augmentation is only applied to the training data. As a result, in some circumstances, a transformation that might be an augmentation may be better as a preprocessing step. Understanding the issue, the data gathered, and the production environment is necessary to determine the preprocessing and augmentation steps for boosting the model's performance.

1. **Resizing**:

   Changing the size of an image sounds trivial, but there are considerations to take into account. Many model architectures call for square input images, but few devices capture perfectly square images. Altering an image to be a square calls for either stretching its dimensions to fit to be a square or keeping its aspect ratio constant and filling in newly created "dead space" with new pixels. Moreover, input images may be various sizes, and some may be smaller than the desired input size.

2. **Tilted 90 degrees to left and right:**

   The rotation operation used for data augmentation is done by randomly rotating the input by 90 degrees zero or more times. Also, horizontal flipping is applied to each of the rotated images

3. **Inverting and Augmentation**:

   Mirroring an image about its x- or y-axis forces our model to recognize that an object need not always be read from left to right or up to down. Flipping may be illogical for order-dependent contexts, like interpreting the text.

Methodology                                                                                                          Chapter 3## 3.3. METHODOLOGY

The overall process starts with obtaining the required datasets, detecting tumors, classifying tumors and then developing an ensemble model

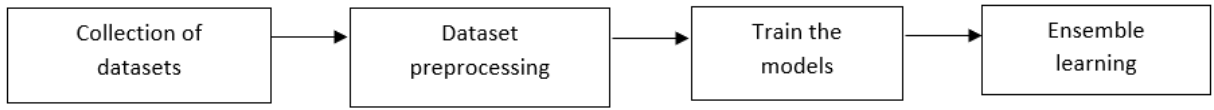

*Fig.3.3. Methodology for detecting brain tumor by ensembling*

Two datasets Brain MRI Images for Brain Tumor Detection and Brain Tumor Classification (MRI) are downloaded from the Kaggle website. The Brain MRI Images for Brain Tumor Detection consists of 253 samples, of which 98 were benign(labeled as 'no') samples while 155 images were malignant(labeled as 'yes') brain tumor samples. The second dataset consists of 3064 images out of which 926 were glioma, 937 were meningioma, 901 were pituitary tumors and 500 were no tumor samples

The aim is to develop an ensemble model that combines the predictions of two models trained with two different datasets respectively and provide an optimal prediction on the samples

There are many machine learning models that are used to predict and classify brain tumors. Four different models were trained for each dataset. Models trained with dataset 1 will predict whether there is a tumor or not. Models trained with dataset 2 will predict the class of a tumor sample.

Then, ensembling is done by combining two models that perform well for the datasets

The implementation is done in Colab Notebook using Python programming language

# CHAPTER 4

# IMPLEMENTATION

In this study, models are used as a deep learning-based feature extractor since they can capture the important features without any human supervision. Here, a total of 8 pre-trained models were used, 4 for dataset 1 and dataset 2.

## 4.1. DenseNet

DenseNet is a type of deep neural network architecture that has shown great success in various computer vision tasks, including image classification. It is a kind of CNN with a number of deeper layers, and has high computational efficiency and storage efficiency, while greatly reducing the number of network parameters. Thus, the doctor combines the MRI images with clinical symptoms and serum biochemical indexes in daily work, which can increase the accuracy of a cancer diagnosis. With the continuous development of science and technology, imaging technology is becoming more and more complex, and treatment plans and results are becoming more and more sophisticated and scientific. In this study, the DenseNet model is used for both Dataset 1 and Dataset 2. It is designed to address the vanishing gradient problem that occurs in traditional CNNs by introducing dense connections between layers. In DenseNet, each layer is connected to all subsequent layers in a feed-forward manner, allowing information to flow directly from the input to the output of the network. This architecture has been shown to improve the performance of CNNs on various image classification tasks. An overview of the algorithm of Densenet is presented in fig. 4.1.

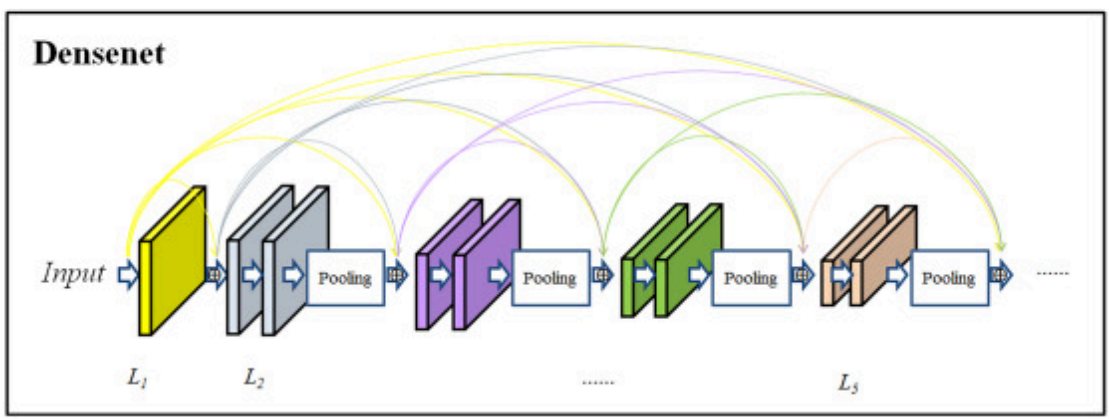

*Fig. 4.1. Layout of DenseNet*

Brain tumor classification is an important medical application of deep learning. DenseNet has been used for brain tumor classification due to its effectiveness in handling medical image data. Researchers have used DenseNet to classify brain tumor images into different categories such as low-grade glioma, high-grade glioma, and meningioma. In a typical brain tumor classification task using DenseNet, the model is trained on a dataset of brain tumor images with known labels. The images are preprocessed to enhance features, and then fed into the DenseNet model for training. During training, the model adjusts its parameters to minimize the difference between its predictions and the true labels of the training data.

The process of brain tumor classification using DenseNet typically involves the following steps:

1. **Data preprocessing:** Medical images, such as Magnetic Resonance Imaging (MRI), are preprocessed to improve image quality and reduce noise. This involves applying filters, normalization, and registration to ensure that the images are in a consistent format.
2. **Feature extraction:** DenseNet is used as a feature extractor to automatically learn and extract meaningful features from the preprocessed images. The output of DenseNet is a set of high-level features that represent different regions of the image, which can be used for classification.
3. **Classification:** The extracted features are fed into a classifier, to classify the brain tumor into different types, such as glioma, meningioma, or pituitary tumor.

Overall, DenseNet has shown great potential in brain tumor classification and is a promising tool for aiding medical professionals in diagnosing and treating brain tumors.

## 4.2. ResNet 50

ResNet-50 is a convolutional neural network architecture that consists of 50 layers, and it has been pre-trained on a large dataset of more than a million images from the ImageNet database. This pretraining enables the network to learn useful feature representations that can be fine-tuned for a variety of computer vision tasks, such as object detection and image classification. In this study, the ResNet model is used for both Dataset 1 and Dataset 2.

ResNet is a deep neural network architecture that uses residual blocks to enable the training of very deep networks with hundreds or even thousands of layers. The residual block consists of two or more convolutional layers with a shortcut connection that bypasses the layers. This shortcut connection adds the output of the convolutional layers to the input of the block, allowing the network to learn residual mappings between layers. An overview of the algorithm of Densenet is presented in fig. 4.2.

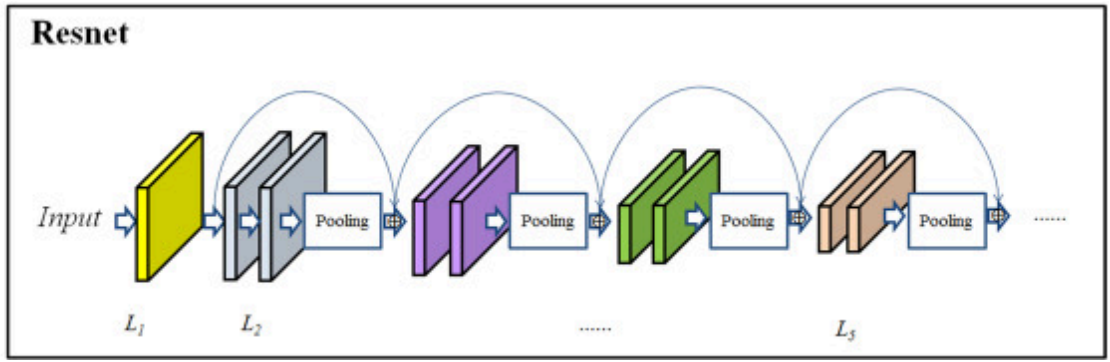

*Fig. 4.2. Layout of ResNet*

The ResNet architecture consists of multiple residual blocks, with each block consisting of two or more convolutional layers with a shortcut connection. The network typically begins with a convolutional layer, followed by a series of residual blocks, and ends with a global average pooling layer and a fully connected layer.

Brain tumor classification is an important medical application of deep learning. It involves the classification of magnetic resonance imaging (MRI) scans of the brain into different tumor categories, such as glioma and meningioma. Deep learning models, including ResNet, have been used for brain tumor classification to achieve high accuracy rates. In a typical brain tumor classification task using ResNet, the model is trained on a dataset of MRI scans with known labels. The images are preprocessed to enhance features, such as by normalizing intensities or removing artifacts, and then fed into the ResNet architecture for training. During training, the model adjusts its parameters to minimize the difference between its predictions and the true labels of the training data. Once the model is trained, it can be used to predict the labels of new brain tumor images. The model takes in an MRI scan as input, processes it through the ResNet network, and outputs a probability distribution over the different categories. The category with the highest probability is chosen as the predicted label.

## 4.3. EfficientNet

EfficientNet is designed to be both computationally efficient and highly accurate, making it well-suited for a range of computer vision tasks, including medical image analysis. Medical image analysis techniques, such as CNNs, have shown great promise in improving the accuracy and efficiency of brain tumor analysis. However, traditional CNN architectures can be computationally expensive and require significant computational resources, limiting their practical utility in clinical settings.EfficientNet addresses this challenge by using a novel compound scaling method to balance model depth, width, and resolution. This allows for the creation of highly efficient CNNs that achieve state-of-the-art performance on a range of computer vision tasks while requiring significantly fewer computational resources. In fact, EfficientNet has been shown to achieve better accuracy than much larger and more computationally expensive models, such as ResNet and Inception, while requiring less than half the computational resources.

For the analysis of brain tumors, EfficientNet, a highly effective and precise CNN architecture, has shown great promise. Its Swish activation function, compound scaling method, and data augmentation techniques make it well-suited for processing large medical image datasets, and numerous studies have shown that it is effective at classifying and segmenting brain tumors. Despite some drawbacks, EfficientNet is a promising advancement in the field of medical image analysis and has the potential to greatly enhance the accuracy and efficiency of brain tumor analysis in clinical settings.

## 4.4. VGG- 16

VGG-16 is a popular and effective model for image classification and has been used in various computer vision applications. VGG-16 is a deep CNN architecture made up of three fully connected layers after the first 16 convolutional layers. By learning filters with progressively more complex properties in the convolutional layers, it was created to extract high-level features from images. The model has a simple and uniform architectural design, which makes it simple to understand and interpret.

When VGG-16 is used to classify brain tumors, a dataset of brain images that have been labeled with the appropriate tumor type is used to train the model. The type of tumor in new brain images can then be predicted using the trained model. The use of VGG-16 in brain tumor classification has several advantages. First, it is a well-established model with a high accuracy rate in image classification tasks. Second, the model has a simple and uniform architecture, which makes it easy to understand and interpret. Third, the model can be trained on a relatively small dataset, which is important in medical applications where collecting large datasets can be challenging.

The training of VGG-16 involves the use of backpropagation algorithm and optimization techniques such as stochastic gradient descent (SGD) to update the weights of the model. The weights of the model are initialized randomly, and the model is trained on a large dataset of labeled images. The objective of training is to minimize the cross-entropy loss function between the predicted probabilities and the true labels of the images. The training of VGG-16 is computationally intensive, requiring powerful GPUs to efficiently train the model. The model is typically trained on a large dataset of images for several epochs, with the number of epochs depending on the size of the dataset and the complexity of the task.

Deep learning-based methods have been shown to improve the accuracy and speed of brain tumor classification. VGG-16 is one of the most widely used CNNs in brain tumor classification, and it has been shown to achieve high accuracy rates. The use of VGG-16 in brain tumor classification involves training the model on a dataset of brain images that have been labeled with their corresponding tumor type. The trained model can then be used to predict the tumor type of new brain images.

VGG-16 is a deep CNN architecture that has been applied to a number of computer vision tasks, including the classification of brain tumors. It has been demonstrated to achieve high accuracy rates when classifying MRI-based brain tumors. The well-proven performance in image classification tasks, the straightforward and uniform architecture, and the capacity to be trained on small datasets are all benefits of using VGG-16 in the classification of brain tumors. VGG-16 is a promising model for upcoming research in the area of medical image analysis because of these benefits.

## 4.5. Vision transformer (ViT)

The Vision Transformer (ViT) is a new architecture that has shown remarkable performance in computer vision tasks, including image classification.

Vision Transformer is a neural network architecture that extends the Transformer model, which was initially introduced for natural language processing (NLP) tasks, to the computer vision domain. The Transformer model consists of an encoder-decoder architecture that uses multi-head self-attention mechanisms to capture the dependencies between different elements of a sequence. The encoder takes the input sequence, and the decoder generates the output sequence.

The architecture of Vision Transformer can be divided into three main components: the patch embedding layer, the Transformer encoder, and the classification head. The patch embedding layer maps the image patches into a high-dimensional feature space, which is then fed into the Transformer encoder. The encoder consists of a series of multi-head self-attention layers and feedforward layers, which learn to capture the dependencies between different patches. The output of the encoder is a sequence of feature vectors, which can be fed into the classification head to obtain the final output.

## 4.6. ENSEMBLE LEARNING

Ensemble learning is a machine learning technique that combines several models to produce one optimal predictive model. This is a method of achieving better predictions by fusing the salient properties of two or more models. The final ensemble model is more robust than the individual models that constitute the ensemble because ensembling reduces the variance in the prediction errors

It is generally used to improve the performance of a model

Ensemble learning is used in various scenarios:

1. Some models may perform better on particular distributions than others. For instance, a model may be well suited to distinguish between cats and dogs but less so when doing so between dogs and wolves. On the other hand, a second model makes inaccurate predictions for the "cat" class while correctly differentiating between dogs and wolves. A combination of these two models could create a decision boundary that is more discriminatory among all three groups of data

2. In situations where there is a large amount of data, instead of training one model with the dataset, the classification jobs can be divided across various classifiers and ensemble them during the prediction period

   In cases where the dataset is small, bootstrapping ensemble technique can be used by creating several subsets of a single dataset using replacement and use the bootstrap samples to train various classifiers

3. Researchers use the confidence scores of the individual classifiers to generate a final confidence score of the ensemble. Involving the confidence scores for developing the ensemble gives more robust predictions than simple "majority voting" since a prediction with 95% confidence is more reliable than a prediction with 51% confidence

4. A problem may occasionally have a complex decision boundary, making it difficult for a single classifier to produce the right boundary. If there is an attempt to solve a problem with a parabolic (polynomial) decision boundary using a linear classifier, for instance. Clearly, a single linear classifier is inadequate for the task. Yet, any polynomial decision boundary can be produced by an ensemble of several linear classifiers

5. Information fusion for improving classification performance is the main justification for employing an ensemble learning model. To acquire a more reliable result, models that have been trained using various data distributions related to the same set of classes are used while making predictions. Integrating the results from these two classifiers will result in more reliable and bias-free predictions

### 4.6.1. METHODS COMMONLY USED IN ENSEMBLE LEARNING

**1. BAGGING**

The Bagging ensemble technique is the acronym for "bootstrap aggregating". The term "bootstrap sampling" refers to the process of creating subsamples from a dataset for this method.ie. replacement is used to generate random subsets of a dataset.

These subsets are handled as separate datasets and several machine learning models will be fitted to them. The predictions from all such models trained on various subsets of the same data are taken into account during test time

The final prediction is calculated using an aggregation process (like averaging, weighted averaging, etc. )

Fig.4.3. explains the bagging ensemble mechanism

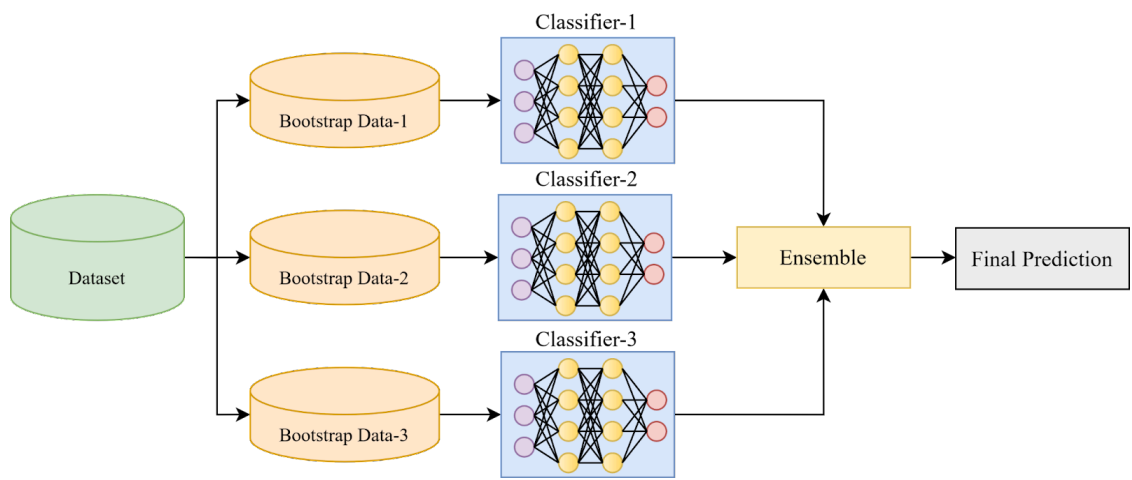

*Fig.4.3. Bagging ensemble mechanism*

## 2. BOOSTING

The dataset is processed in sequential order. The complete dataset is supplied to the first classifier, and the predictions are analyzed. The samples close to the decision boundary of the feature space, where Classifier-1 failed to make accurate predictions, are sent to the second classifier.

This is done so that Classifier-2 can pay close attention to the feature space's problematic regions and learn the proper decision boundary. Similar to this, more steps of the same concept are used, and the final prediction on the test data is derived using the ensemble of all these prior classifiers

In general, the classifiers selected for the ensemble must have high bias and low variance, i.e., simpler models with fewer trainable parameters

Fig.4.4.  explains Boosting ensemble mechanism

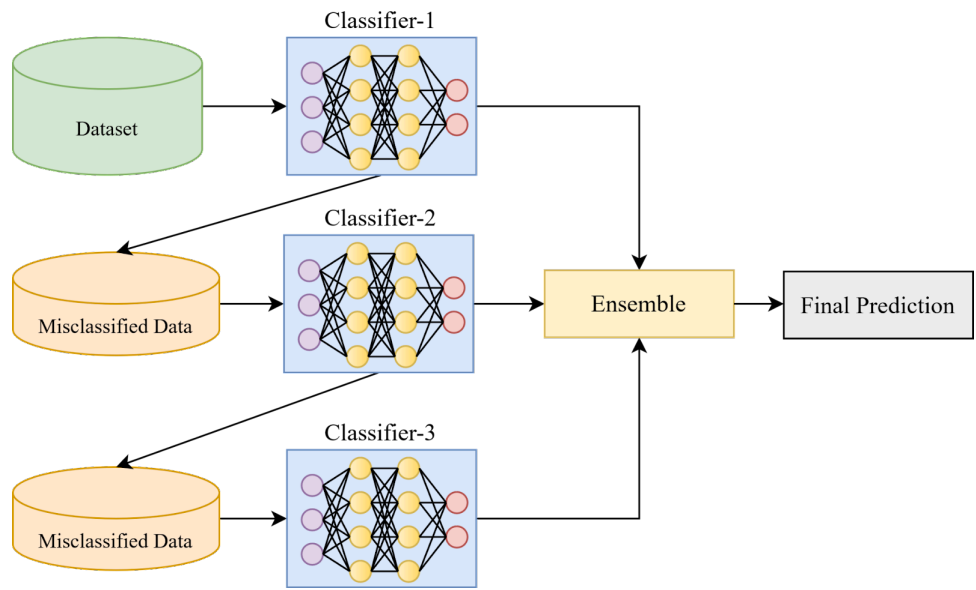

*Fig.4.4.  Boosting ensemble mechanism*

## 3. STACKING

Similar to the bagging ensemble process for training multiple models, the stacking ensemble method also involves the creation of bootstrapped data subsets.

The results of all such models are fed into a meta-classifier, a different classifier that ultimately predicts the samples.  Fig.4.5. explains Stacking ensemble mechanism

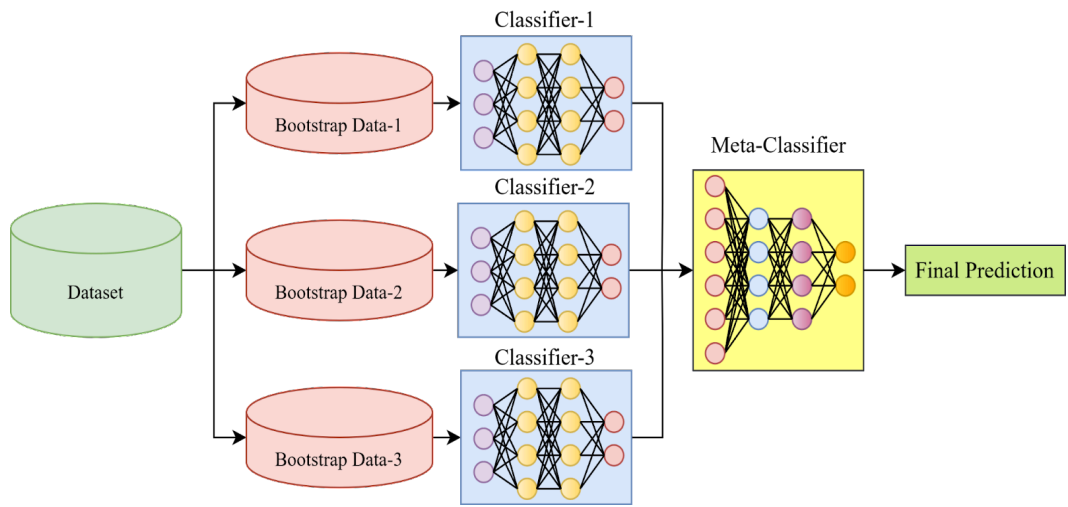

*Fig.4.5.  Stacking ensemble mechanism*

## 4. MIXTURE OF EXPERTS

The "Mixture of Experts" ensemble genre trains a number of classifiers, combining their results according to a generalized linear rule.

A "Gating Network," another trainable model—typically a neural network—also determines the weights assigned to these pairings.

When various classifiers are trained on separate regions of the feature space, an ensemble technique like this one is typically utilized

 Fig.4.6. explains Mixture of Experts ensemble mechanism

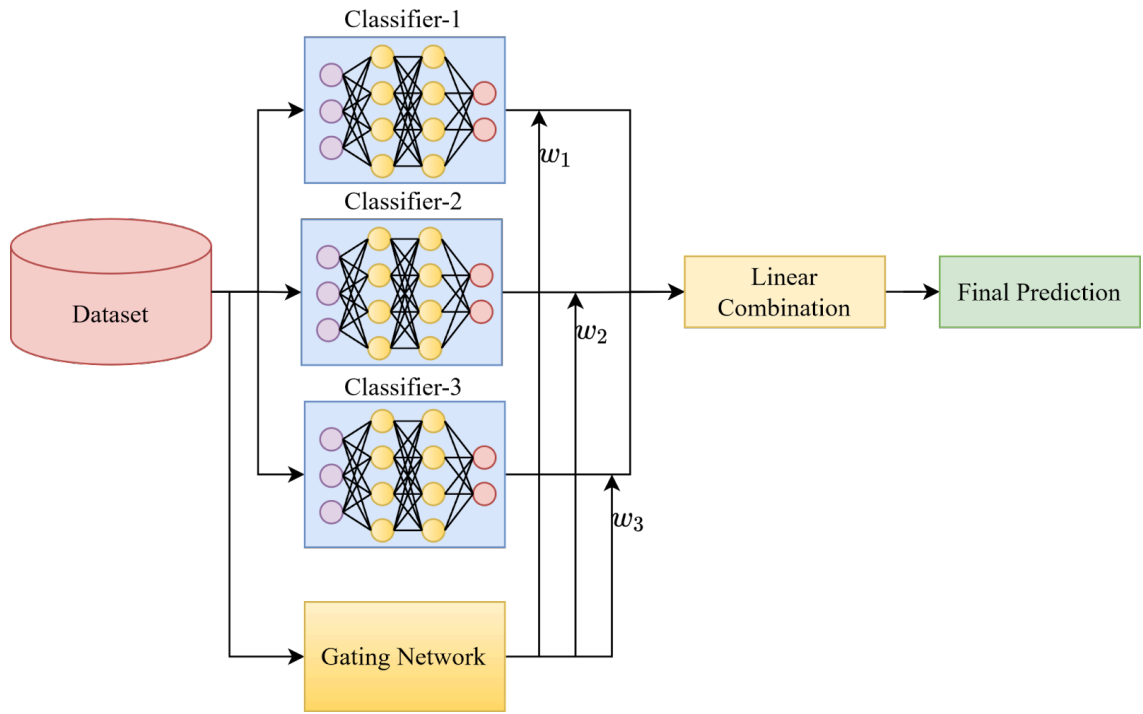

*Fig.4.6. Mixture of Experts ensemble mechanism*

## 5. MAJORITY VOTING

This approach selects an odd number of contributing classifiers, and then computes the classifiers' predictions for each sample. The class that draws the greatest amount of data from the classifier pool is then known as the ensemble's anticipated class

The class that gets most of the class from the classifier pool is deemed the ensemble's predicted class. This method works well for binary classification problems but fails for a problem with many classes

In these circumstances, a random class from the top candidates is selected, which results in a larger margin of error. As a result, techniques based on confidence scores are now more often employed and more trustworthy

## 6. MAX RULE

The probability distributions that each classifier produces are used as the foundation of the "Max Rule" ensemble approach. This approach uses the idea of the classifiers' "confidence in prediction," making it a better approach for multi-class classification problems than majority voting. Here, the associated confidence score for a class predicted by a classifier is examined. The ensemble framework's prediction is the class prediction made by the classifier with the highest confidence score

## 7. PROBABILITY AVERAGING

The probability scores for several models are initially computed in this ensemble technique. The results are then averaged across all models for all classes in the dataset

Probability scores represent a model's level of assurance in its predictions. In order to produce a final probability score for the ensemble, the confidences of various models are pooled in this instance. The anticipated class is designated as the one with the highest probability following the averaging operation

## 8. WEIGHTED PROBABILITY AVERAGING

In this method, the weighted average of the probability is calculated. The weights in this approach refer to the importance of each classifier, i.e., A classifier whose overall performance on the dataset is better than another classifier is given more importance while computing the ensemble, which leads to a better predictive ability of the ensemble framework.

Recent methods focus on leveraging weights. For datasets where each class contains a different quantity of training data, assigning the classification accuracy as the weights to the ensemble can aggravate the performance

# CHAPTER 5

# RESULT AND DISCUSSION

The results were obtained for two different datasets (dataset 1 and dataset 2) for the tasks of the brain tumor classification.

The datasets are preprocessed and are split into training and testing dataset files where 80% of the data is used for training and the rest for testing.

For Brain Tumor MRI Images for dataset 1, four models were trained. The trained models are DenseNet, ResNet, EfficientNet and VGG-16. The accuracies of the models are presented in table 5.1

*Table 5.1. Models trained with Brain Tumor MRI Images for dataset 1 and their accuracies*

| MODEL | ACCURACY(%) |
|---|---|
| DENSENET | 71.43 |
| RESNET | 80.36 |
| EFFICIENT NET | 66 |
| VGG-16 | 80.63 |

For Dataset 2, four models were trained. The models and their accuracies are presented in the table

*Table 5.2. Models trained with dataset 2 and their accuracies*

| MODEL | ACCURACY(%) |
|---|---|
| DENSENET | 84.32 |
| RESNET | 50 |
| EFFICIENT NET | 77 |
| VISION TRANSFORMER | 81 |

From the above tables the models with higher accuracies are chosen. Then, an ensemble model is developed by combining two models with higher accuracies from tables 5.1 and 5.2 respectively. The ensemble accuracy of 91.07% is obtained

# CHAPTER 6

# CONCLUSION AND FUTURE WORK

In conclusion, the development of a machine learning model for the classification of primary stage brain tumors has significant potential in improving the accuracy and speed of diagnosis. The project utilized a dataset of MRI scans to train and test the machine learning model, achieving a high level of accuracy in distinguishing between different types of primary brain tumors.

The use of machine learning in medical imaging analysis can provide significant benefits, including improved accuracy, speed, and consistency in diagnosis. It also has the potential to assist healthcare professionals in making more informed decisions regarding patient treatment plans.

The project's findings have important implications in the field of medical imaging analysis and the broader healthcare industry. With further development and refinement, machine learning models can provide a valuable tool for the early detection and diagnosis of brain tumors, potentially leading to better patient outcomes and improved survival rates.

Overall, the successful development and implementation of a machine learning model for the classification of primary stage brain tumors is a significant step forward in the fight against brain cancer and demonstrates the potential for machine learning to revolutionize medical diagnosis and treatment.

The next phase of this project includes incorporating more types of tumor especially secondary brain tumors. This can be acheived by adding more images to the data set and ensembling more heavy models to increase the accuracy and precision of classification. Further, after scaling  the model there is a scope for automation and with the help of healthcare professionals who can integrate the machine learning model into clinical practice would help to validate its effectiveness in real-world scenarios.